\documentclass{elsart5p}

\usepackage{graphics}
\usepackage{graphicx}
\usepackage{amssymb}
\usepackage{bm}

\begin{document}

\begin{frontmatter}



\title{Intermediate scale phenomena in cuprates}
%

\author[AA]{J. Lorenzana\corauthref{Name1}},
\ead{jose.lorenzana@roma1.infn.it}
\author[BB]{G. Seibold}

\address[AA]{SMC-INFM,ISC-CNR, Dipartimento di Fisica,
Universit\`a di Roma ``La Sapienza'', P. Aldo Moro 2, 00185 Roma, Italy }
\address[BB]{Institut f\"ur Physik, BTU Cottbus, PBox 101344,
         03013 Cottbus, Germany}

\corauth[Name1]{Corresponding author. Tel: (39) 0649937511 fax: (39) 0649937440
}

\begin{abstract}
We present computations of cuprates which
accurately describe the physics arising at scales of up to several lattice
constants. 
\end{abstract}


\end{frontmatter}


After 20 years of research the low energy physics of cuprates remains
a mystery. It is generally accepted that a three band Hamiltonian  
with Cu's and O's orbitals should contain the essential physics so
we can say that there is consensus at the
length scale of the Cu-O distance but there is disagreement when it
comes to describe higher length scales or to set up a low energy model.

Much on the modeling in the field consists on guessing the low energy
physics. 
 We adopt
a more conservative approach and try to move from the safe land of the
atomic scale to the intermediate scale of a few atomic constants. 
An exact solution is not possible but a number of 
techniques allow to address the intermediate length scale physics
with accuracy. Rather than guessing we can accurately compute what
the model predicts at intermediate energies and length scales and
compare in detail with experiments. This should considerably constrain
the low energy physics. 
If our modeling is successful we should be able to describe
spectroscopies from high energies up to some infrared cutoff (below 
which our ignorance sets in) and  we should be able to
describe short and intermediate length scale physics. This includes the
ground state energy which, in a short range model, is essentially a short range property. 


In the last years we have carried  on this program pushing in some cases
our 
``discerning cutoff''  down to a few meV. We have based our
computations on the Gutzwiller approximation and a time dependent
extension\cite{sei01}. These techniques have revealed to be very
accurate when compared with exact
diagonalization\cite{sei01,sei03,sei04b} and exact results in 
one\cite{sei01} and infinite\cite{gun07} dimensions. 
An accurate
technique supplemented with an accurate model should produce results in accord with
experiment and it does.

Within our approximations the lowest energy
solutions consist of O centered stripes which have approximately a
number of holes per lattice site along the stripe $\nu\sim 0.5$. These
textures  explain  the behavior of the
incommensurability, the chemical 
potential, some anomalous transport experiments\cite{lor02b} 
and the  optical
conductivity as a function of doping\cite{lor03}. 
 \begin{figure}[b]
 \includegraphics[width=7cm,clip=true]{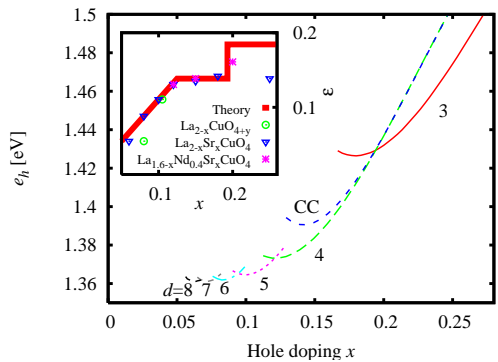}%
 \caption{$e_h$ as a function of doping for vertical
O centered stripes.
 We also show the result
    for Cu centered (CC) stripes   
for $d=4$. The inset in 
reports the incommensurability as obtained from the present calculation
(line) compared with experimental data from Ref.~\protect\cite{yam98}.}
 \label{fig:ehdx}
 \end{figure}
In order to get accurate parameters we have
relied on LDA.  
New calculations in much larger
systems which allow better resolution show  small deviation from
experiment. For example the filling of the stripes with the parameters
of Ref.~\cite{lor02b} is $\nu=0.56$ holes per site
 rather than  $\nu=0.5$ as found in experiment\cite{yam98}. This
 can be easily understand from the fact that the stripe filling is
 approximately given by $\nu\sim\sqrt{J/2C}$ with $C$ the inverse
 compressibility of the stripe and $J$ the superexchange interaction\cite{sei04a,lor02b}. 
LDA parameters are well known to overestimate the value of $J$ so it
is not surprising that the filling of the stripe is overestimated. To
improve the agreement with experiment here we present new results with
an empirical parameter set based on Ref.~\cite{esk90}. 
 We use $\Delta=3.3$eV for Cu-O onsite energy splitting,
 $t_{pd}=1.3$eV ($t_{pp}=0.7$eV) for Cu-O (O-O) hopping,
 $U_{d}=8.8$ eV ($U_{p}=6.0$ eV) for onsite
repulsion on Cu (O), $U_{pd}=1.0$ eV for Cu-O
repulsion and $K_{pd}=-0.25$ for Cu-O direct exchange. The CuO
interactions $U_{pd}$ and $K_{pd}$ where added by us. $K_{pd}$ has been recognized as
essential to yield the correct 
value of $J$\cite{ste88} and so it plays an unsuspected role on fixing
the filling of the stripe. 
We have found that  the optical conductivity and the magnetic
excitations computed within the time dependent Gutzwiller approximation\cite{sei01} 
for the undoped system are in excellent agreement with the
experiment confirming the accuracy of the parameters. 

Fig.~\ref{fig:ehdx} shows the energy per
hole as a function of doping. Results are qualitatively similar as in
Ref.~\cite{lor02b} but now the stripes have $\nu\sim1/2$ even for large systems.
As discussed before\cite{lor02b} for doping $x\lesssim 1/8$ the system responds to a
change of doping by changing the periodicity $d$ whereas for
$x\gtrsim 1/8$ the periodicity gets fixed at $d=4$ in a large doping
range. The inset shows the resulting magnetic 
incommensurability compared  with experiment.

\begin{figure}[b]
\begin{center}
\includegraphics[angle=0,width=0.45\textwidth]{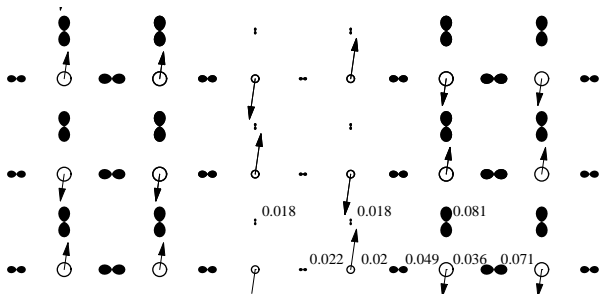}
\caption{ Open circles ($p$ orbitals) represent Cu (O) sites. The
  numbers and the size of symbols  represent the excess charge  whereas the
spin density is proportional to the length of arrows.   
}
\label{rhosp}
\end{center}
\end{figure}

Fig.~\ref{rhosp} shows the charge and spin density at doping
$x=1/8$. A similar charge distribution was predicted in Ref.~\cite{lor02b} and
found to be in excellent agreement with a charge sensitive probe by
Abbamonte an collaborators\cite{abb05}. They could not determine if the stripes where Cu centered
or O centered. More recently Davis
and collaborators\cite{koh07} have imaged glassy stripes which indeed are centered
on O as  predicted\cite{lor02b}.  Taken together these experiments
give us amplitude and phase information of stripes in excellent
agreement with Ref.~\cite{lor02b} thus showing that it is possible to obtain 
and even predict realistic information on the intermediate scale 
physics of cuprates. 

\begin{figure}[t]
\begin{center}
\includegraphics[angle=0,width=0.45\textwidth]{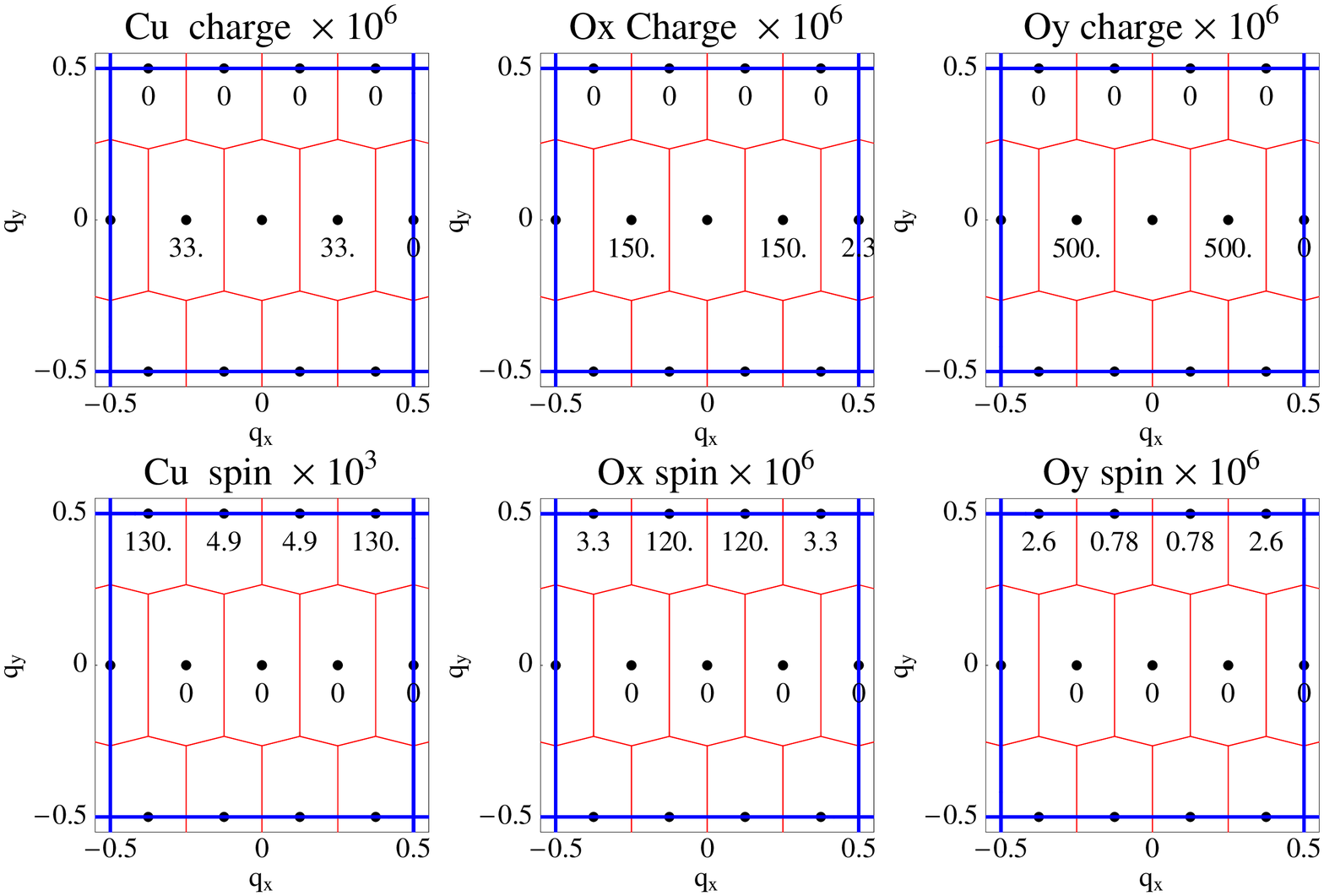}
\caption{Bragg weights $m_{\bm{Q}}^2$, $n_{\bm{Q}}^2$ at reciprocal
  lattice vectors (indicated by the dots) for $x=1/8$ and $d=4$
  stripes. The polygons around each
  point are the reduced magnetic Brillouin zones. We use reciprocal
  lattice units of the fundamental Brillouin zone. Ox (Oy) are the
  oxygens in the bonds perpendicular (parallel) to the stripe. 
}\label{bragg}
\end{center}
\end{figure}

The Fourier transform of the charge and spin distribution determines
Bragg peak weights ($\propto m_{\bm{Q}}^2$) in scattering experiments
(for the definitions see Ref.~\cite{lor05}) as shown in
Fig.~\ref{bragg}.  Disregarding the 
difference in cross section for different processes (magnetic neutron scattering
vs. X-ray or nuclear neutron scattering) 
we see that Bragg weights are practically 3 orders of
magnitude smaller for charge than for spin. This is due to very soft
charge distribution shown in Fig.~\ref{rhosp} with respect to the spin
distribution and explains why it has been so hard to detect  charge
ordering.  

Optical excitation on top of these textures and 
magnetic excitations on top of similar textures in the one band
Hubbard model have been found to be in excellent agreement with
experiment confirming the validity of the intermediate scale physics found\cite{sei05,sei06}.

To conclude our computations are based on a mean-field like approach and therefore
can not address subtle issues as the absence or presence of long
range order, specially in the delicate magnetic channel, but we can
reliably determine the behavior at intermediate length scales
improving our understanding of these fascinating materials. 
Our results provide strong constrains on low energy theories.

\end{document}